\documentclass[letterpaper]{article} 
\usepackage{aaai25}  
\usepackage{times}  
\usepackage{helvet}  
\usepackage{courier}  
\usepackage[hyphens]{url}  
\usepackage{graphicx} 
\usepackage{natbib}  
\usepackage{caption} 
\usepackage{algorithm}
\usepackage{algorithmic}
\usepackage{times}
\usepackage{epsfig}
\usepackage{graphicx}

\usepackage{amssymb}
\usepackage{caption}
\usepackage{bbding}
\usepackage{multirow}
\usepackage{subcaption}
\usepackage{tabularx}
\usepackage{booktabs}
\usepackage[autostyle=true]{csquotes}
\usepackage[dvipsnames]{xcolor}
\usepackage{amsmath}
\usepackage{mathtools}
\usepackage[dvipsnames]{xcolor}
\usepackage{xspace}
\usepackage[dvipsnames]{xcolor}

\newcommand{\CUT}[1]{}

\usepackage{mathtools}

\usepackage{newfloat}
\usepackage{listings}
\DeclareCaptionStyle{ruled}{labelfont=normalfont,labelsep=colon,strut=off} 
\floatstyle{ruled}
\newfloat{listing}{tb}{lst}{}
\floatname{listing}{Listing}
\title{CAMH: Advancing Model Hijacking Attack in Machine Learning}
\author{
    Xing He\textsuperscript{\rm 1}\equalcontrib, 
    Jiahao Chen\textsuperscript{\rm 1}\equalcontrib, Yuwen Pu\textsuperscript{\rm 1}, Qingming Li\textsuperscript{\rm 1}\thanks{Corresponding author}, Chunyi Zhou\textsuperscript{\rm 1},\\ Yingcai Wu\textsuperscript{\rm 1}, Jinbao Li\textsuperscript{\rm 23}, Shouling Ji\textsuperscript{\rm 1}
}
\affiliations{
    \textsuperscript{\rm 1}College of Computer Science and Technology, Zhejiang University\\
    \textsuperscript{\rm 2}Shandong Artificial Intelligence Institute\\
    \textsuperscript{\rm 3}School of Mathematics and Statistics, Qilu University of Technology\\

}

\usepackage{bibentry}

\begin{document}

\maketitle

\begin{abstract}

In the burgeoning domain of machine learning, the reliance on third-party services for model training and the adoption of pre-trained models have surged. However, this reliance introduces vulnerabilities to model hijacking attacks, where adversaries manipulate models to perform unintended tasks, leading to significant security and ethical concerns, like turning an ordinary image classifier into a tool for detecting faces in pornographic content, all without the model owner's knowledge. This paper introduces Category-Agnostic Model Hijacking (CAMH), a novel model hijacking attack method capable of addressing the challenges of \textit{class number mismatch}, \textit{data distribution divergence}, and \textit{performance balance between the original and hijacking tasks}. CAMH incorporates synchronized training layers, random noise optimization, and a dual-loop optimization approach to ensure minimal impact on the original task's performance while effectively executing the hijacking task. We evaluate CAMH across multiple benchmark datasets and network architectures, demonstrating its potent attack effectiveness while ensuring minimal degradation in the performance of the original task.Our Code  \url{https://github.com/healthandAI/CAMH}
\end{abstract}

\section{Introduction}
In today's digital era, machine learning (ML) models have become essential tools in critical fields such as finance, healthcare, and autonomous driving \cite{li2019target,zhang2020unsupervised,liu2020re}. Training machine learning models is expensive, as it requires costly computational resources and precise hyperparameter tuning by domain experts \cite{expensive}. To reduce these costs, organizations are increasingly turning to third-party services like Google AutoML \cite{GoogleCloud}, Amazon SageMaker \cite{Amazonsagemaker}, or Microsoft Azure ML \cite{MicrosoftAzure} for neural network training or downloading pre-trained models from repositories such as Hugging Face \cite{Huggingface} or Model Zoo, especially those trained on popular datasets \cite{ModelZoo}.

However, this practice exposes models to real-world security risks, particularly model hijacking attacks. Model hijacking attacks enable adversaries to take control of a target model and compel it to perform completely different tasks than its original one. Successfully executing a model hijacking attack can lead to two principal risks: accountability risk and parasitic computing risk  \cite{salem2022get}. 

The accountability risk means that attackers could use the model to conduct illegal or unethical tasks, such as turning an ordinary image classifier into a tool for identifying faces in pornographic content, all without the model owner's knowledge. This could lead to the model owner being wrongly accused of providing illegal services despite not being directly involved. Additionally, the owners of hijacked models face the risk of parasitic computing, where attackers exploit the hijacked model to run their own applications without incurring the costs of operating, and maintaining the models themselves, which can be prohibitively expensive in some regions (see detailed scenarios in Section \ref{case1}).

Salem et al. \cite{salem2022get} first introduced the concept of model hijacking attacks. They demonstrated during training how a target model could be redirected, such as transforming an image classifier into a facial recognition tool. These attacks degrade model performance and expose model owners to legal and ethical risks. Si et al. \cite{si2023two} further expanded this research domain by introducing new attack strategies applicable to complex scenarios beyond image classification, including data manipulation and task transfer. The hijacking attacks are an emerging threat to model security, and the only two current methods primarily focus on modifying poisoning operations within the training dataset, which means that the model’s output remains constrained to the number of categories defined by the original task. Consequently, these methods impose significant restrictions on the flexibility of hijacking tasks, such as requiring the number of categories in the hijacking task not to exceed those in the original task and ensuring that the data distribution of the original task does not differ significantly. If these conditions are not met, it becomes impossible to balance the hijacking task and the original task's performance, resulting in loss of concealment or hijacking failure.

We hope that model hijacking attacks will no longer be limited by previous methods. In the implementation process, we summarize three major challenges:

\begin{itemize}

\item \textbf{C1}: \textit{How do we solve the limitation that the number of hijacking task categories cannot exceed the number of original task categories?}
\item \textbf{C2}: \textit{How to ensure the effectiveness of model hijacking attacks given different data distributions?}
\item \textbf{C3}: \textit{How do we balance the performance of the original task with the execution of the hijacking task?}
\end{itemize}

To address these challenges, we propose a novel model hijacking attack method—CAMH. We've integrated synchronized training layers to handle different class numbers in model training. We've also optimized noise to adjust for data distribution variations. Finally, we use a dual-loop approach to keep the original task's performance stable. To assess the effectiveness of the CAMH method comprehensively, we have established two evaluation metrics: Camouflage Ratio (CR) and Exploitability Ratio (ER). The Camouflage Ratio gauges the performance discrepancy between the hijacked model and the benign model on the original task, whereas the Exploitability Ratio measures the performance gap between the hijacked model and the benign model on the hijacking task. We evaluate the effectiveness of CAMH on multiple benchmark tasks (including MNIST, SVHN, GTSRB, CIFAR10, and CIFAR100) across various network architectures (including ResNet18 and ResNet34). The experimental results demonstrate that CAMH successfully executes hijacking tasks, with the ER values exceeding 85\% for most datasets and the CR maintained at approximately 98\% for all datasets.

The main contributions of this paper include:

\begin{itemize}

    \item We propose a novel model hijacking attack method, CAMH, which, for the first time, overcomes the limitation that the number of categories in the hijacked dataset cannot exceed those in the original dataset.
    \item We proposed synchronized training layers, random noise optimization, and a dual-loop optimization approach to address challenges such as class number mismatch, data distribution divergence, and balancing performance between the original and hijacked tasks. 

    \item We validated CAMH's effectiveness through evaluations on various benchmark datasets and network architectures. Even when the hijacking dataset's categories were four times greater or the hijacking data volume was as low as 30\%, the ER exceeded 85\%, confirming CAMH's robustness.

\end{itemize}

\section{Related Work}

In this section, we review some of the related works. We divide the related works into backdoor attacks in machine learning and model hijacking attacks. Backdoor attacks can be seen as a specific instance of model hijacking attacks. The research into backdoor attacks helps us to deeply comprehend the intrinsic mechanisms of model manipulation and provides a basis for mastering more widespread strategies for model hijacking. We start with the backdoor attacks, followed by model hijacking attacks.

\subsection{Backdoor Attacks in Machine Learning}

Backdoor attacks pose a significant security threat in the field of machine learning, where the adversary manipulates the target model’s training to backdoor it \cite{backdoorsurvey}. The backdooring behavior is usually assigned with a trigger, which is when inserted in any input sample, the target model predicts a specified label. Gu et al. \cite{gu2017badnets} introduced BadNets, the first backdoor attack against machine learning. BadNets uses a white square at the corner of the images as a trigger to misclassify the backdoored inputs to a specific label. Salem et al. \cite{salem2022dynamic} later proposed a dynamic backdoor, where instead of using a fixed trigger, they used a dynamic one. Another similar attack is the Trojan attack \cite{liu2018trojaning,trojan2}. This attack simplifies the assumptions of the backdoor attack by not assuming the knowledge of any sample from the distribution of the target model’s training dataset. There also exist multiple backdoor attacks against Natural Language Processing (NLP) models \cite{chen2021badnlnlp}, federated learning \cite{fl}, video recognition \cite{zhao2020clean}, transfer Learning \cite{yao2019latent}, diffusion model \cite{ho2020denoising}, Chatgpt \cite{shi2023badgpt} and others \cite{saha2020hidden,rakin2020tbt,liu2020reflection,li2021invisible,hayase2022few}. 

The backdoor attack can be considered a specific instance of the model hijacking attack by classifying the backdoored samples as the hijacking dataset. However, our model hijacking attack is more general, i.e., it poisons the model to implement a completely different task.

\subsection{Model Hijacking Attack}
Model hijacking attacks, in which attackers attempt to manipulate a target model to perform unintended tasks, are an emerging attack that can lead to serious security and liability issues. Salem et al. \cite{salem2022get} introduced the concept of model hijacking attacks within the realm of computer vision tasks. Their work is similar to SSBA \cite{li2021invisible} have proposed techniques to enhance the stealthiness of backdoor attacks, thereby reducing the effectiveness of backdoor detection mechanisms. Expanding the scope of model hijacking attacks, Si et al. \cite{si2023two} introduced a novel attack named Ditto, which extends this threat to text generation and classification models.

While both Salem et al. and Si et al. have significantly advanced the understanding of model hijacking attacks, their work exhibits several limitations. First, the effectiveness of these attacks is highly contingent upon the adversary’s capability to generate stealthy modifications to the training data, which may be detectable by robust data monitoring and anomaly detection systems. Second, their approaches impose relatively stringent constraints on the hijacking dataset. For instance, the number of categories in the hijacking dataset must not exceed that of the original dataset. Third, they do not address emerging scenarios such as outsourcing and model marketplaces.

In our study, we thoroughly introduce CAMH, a innovative model hijacking attack designed for new scenarios. CAMH demonstrates exceptional attack performance even under constraints of multi-class hijacking and significant data discrepancies while avoiding the need for deceptive modifications to the training dataset.

\section{CAMH}
In this section, we present various techniques for model hijacking attacks. We first introduce the threat model and two attack scenarios. Then, we propose three specific CAMH techniques: Multi-Class Hijacking Task Design, Noise Optimization Technique, and Dual-loop Optimization.

\subsection{Threat Model and Attack Scenario}
\subsubsection{Threat Model.}Model hijacking attacks can be widely applicable to any real-world scenario where model owners can independently train models, such as outsourcing, model marketplaces, federated learning \cite{bonawitz2019towards}, etc. In the attack scenario proposed in this study, adversaries have complete control over the model training process, and obtain the logits output by the model. We assume adversaries possess a hijacked dataset, which they use to covertly embed the model with hijacking tasks. Whether through outsourced model training or adversaries independently deploying malicious models in the model marketplace, adversaries can access the training dataset, with the distinction being that the former is provided by the victim company and the latter is owned by the adversaries themselves. Additionally, attackers do not need to create disguised datasets as in previous work, nor do hijacked datasets need to resemble the visual features of the original dataset.
\begin{figure}[t]
\centering
\includegraphics[width=0.9\columnwidth]{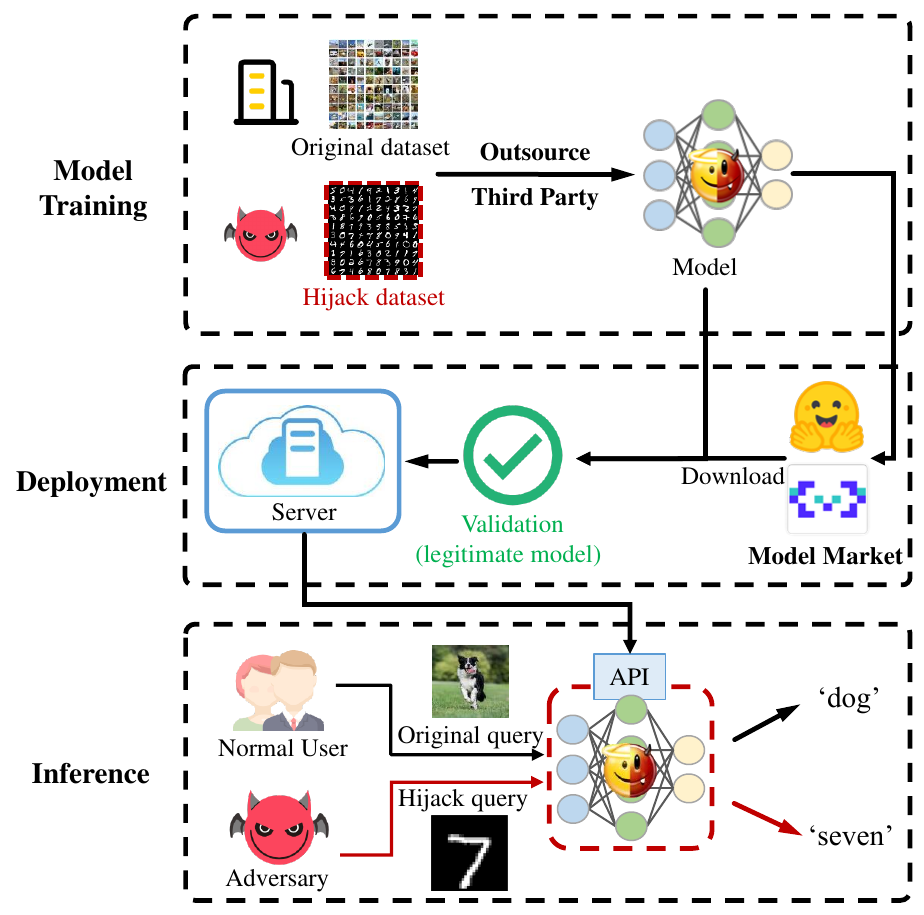} 
\caption{CAMH attacks explored in this paper.}
\label{fig1}
\end{figure}

As shown in Figure \ref{fig1}, we show two attack scenarios, namely outsourcing and model market, which we will explain in detail below.
\subsubsection{Model Hijacking in Outsourcing Scenarios.}
\label{case1}
 In the current field of machine learning, model companies often outsource model training tasks to external entities \cite{outsourced}, which may possess unique data resources or computational capabilities. While this collaborative model can bring technical advantages and cost-effectiveness, it also introduces potential security risks. The attack scenario discussed in this paper involves the outsourcing training process, where if the training party is malicious, they can leverage our model hijacking attack method to clandestinely implant an additional hijacking task in the model while completing the tasks assigned by the model company.  

Specifically, the malicious training party would cleverly incorporate their own hijacking data alongside the data provided by the model company during the training process. These hijacking data are carefully crafted to have a negligible impact on the model's performance for its original task, yet they are sufficient for the model to learn to execute a completely different hijacking task. Once the model has been trained and is publicly released by the model company via APIs \cite{tramer2016stealing}, attackers can execute the hijacking task within the model through these API calls, leading it to perform actions that are illegal or unethical.

For example, attackers could hijack an image classification model to perform sensitive tasks such as facial recognition or sentiment analysis when receiving specific inputs, tasks that are illegal without explicit authorization. This not only leads to wastage of computational resources and maintenance costs for the model company but also exposes them to legal liability and reputational damage due to the model's inappropriate 
behavior.

\subsubsection{Covert Hijacking in Model Marketplaces.}
\label{case2}
In the field of machine learning, model marketplaces such as Hugging Face \cite{Huggingface} or the Model Scope \cite{ModelScope} have become crucial channels for sharing and distributing pre-trained models. These platforms offer users abundant resources, enabling them to rapidly deploy state-of-the-art machine learning models without needing to train from scratch. However, this convenience also brings new security risks \cite{modelmarket,modelmarket2}.

In our envisioned attack scenario, attackers exploit these model marketplaces as distribution channels to propagate models with covert hijacking tasks. Attackers first train a model through standard procedures to perform a seemingly useful original task (e.g., image classification or language translation) while training with the hijacking dataset, the hijacking task is stealthily incorporated. Subsequently, attackers package this model with the covert hijacking task as one solely possessing original task functionality and upload it to model marketplaces for distribution.

During promotion, attackers deliberately conceal the hijacking nature of the model, emphasizing only its performance and advantages in the original task. Once other users or companies download and deploy this model, especially when they provide it with API access, attackers can utilize this API to remotely activate and execute the hidden hijacking task embedded within the model. This covert hijacking task could involve illegal or unethical activities, such as unauthorized data collection, sensitive information classification, or other malicious behaviors.

This form of attack not only poses risks to the model's end-users but also has the potential to severely damage the reputation and trustworthiness of model marketplaces. Users may face legal liabilities and ethical reproach for employing seemingly legitimate but actually hijacked models.

\subsection{Design of CAMH}
As highlighted in Section 1, the realm of model hijacking confronts three principal challenges, which can be succinctly encapsulated as the mismatch in class numbers, divergence in data distributions, and the delicate balance of performance. In direct response to these intricacies, we have engineered a novel assault strategy aimed at accomplishing model hijacking. This section delineates the intricate design considerations that underpin our approach.

\subsubsection{Multi-Class Hijacking Task Design.}
 To address the limitation that the number of hijacking task categories cannot exceed the number of original task categories (C1), our CAMH approach includes an innovative Multi-Class Hijacking Task Design. This design introduces a Synchronized Optimization Layer (SOL) within the model architecture. Positioned after the output classification layer, the SOL is a fully connected layer that adeptly transforms the output logits from the original classification task into a space that aligns with the hijacking task's output dimensions. By leveraging this layer, our method gracefully navigates the complexities arising from an imbalanced number of classes, enabling the seamless integration of a hijacking task into the model's learning process. This layer is specifically designed to manage the data mapping for the hijacking task while leaving the data for the original task unaffected by its processing. This strategic design ensures that the performance of the original task remains stable and uncompromised.
 
Specifically, the mathematical expression of the SOL can be described as follows:
\begin{equation}
\centering
\begin{aligned}
    & f_{\theta}(x):x\in\mathbb{R}^{B\times C\times H\times W}{\longrightarrow}y\in\mathbb{R}^{B\times C_{1}} \\
    & \mathit{h}_{\hat{\theta}}(y):y\in\mathbb{R}^{B\times C_{1}}{\longrightarrow}z\in\mathbb{R}^{B\times C_{2}}
\end{aligned}
\end{equation}

Here, $C_{1}$ and $C_{2}$ represent the class cardinalities of the original and hijacking datasets, respectively. The mapping function 
$f_{\theta}$ represents the model's internal mechanism for processing the input data $x$, which are the input features or images the model receives, whereas $h$ delineates the intermediate mapping within a layer. The term $B$ signifies the batch size, with $C$, $H$, and $W$ representing the channel count, height, and width of the image matrix, respectively.  The output $y$ is the model's prediction for the original task, providing a set of class probabilities or scores based on $x$. Finally, $z$ corresponds to the model's output for the hijacking task after being processed by the SOL.

\begin{figure}[t]
\centering
\begin{minipage}[t]{0.24\linewidth} 
\centering
\includegraphics[width=\linewidth, height=\linewidth]{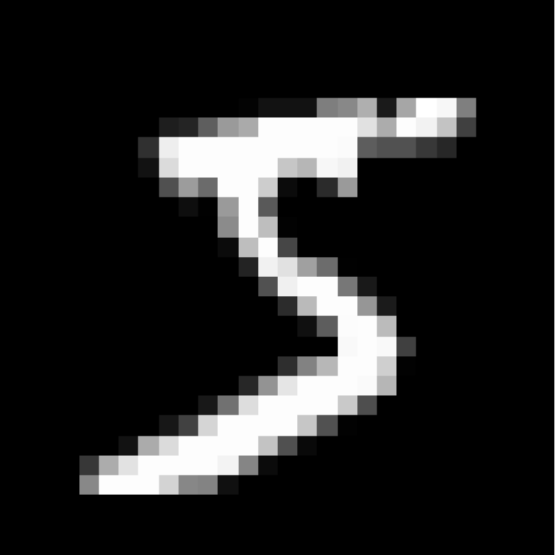} 
\caption*{(a)}
\label{fig4a}
\end{minipage}
\begin{minipage}[t]{0.24\linewidth}
\centering
\includegraphics[width=\linewidth, height=\linewidth]{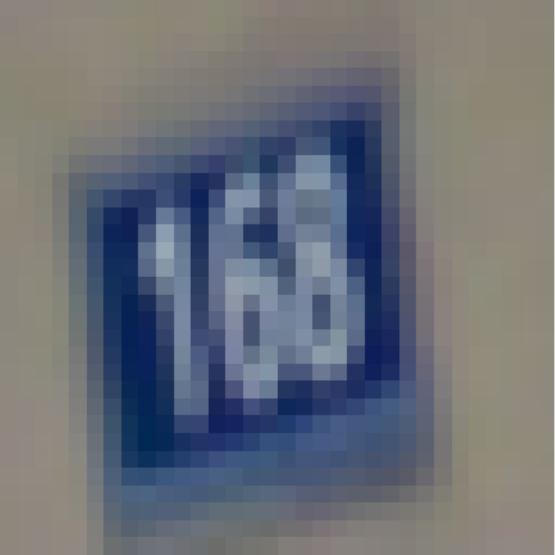}
\caption*{(b)}
\label{fig4b}
\end{minipage}
\begin{minipage}[t]{0.24\linewidth}
\centering
\includegraphics[width=\linewidth, height=\linewidth]{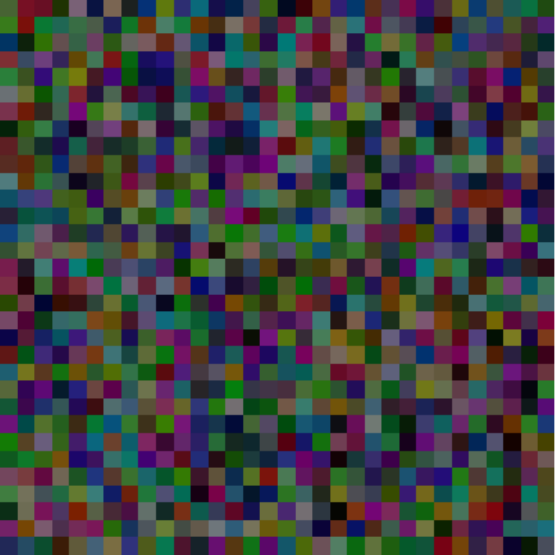}
\caption*{(c)}
\label{fig4c}
\end{minipage}
\begin{minipage}[t]{0.24\linewidth}
\centering
\includegraphics[width=\linewidth, height=\linewidth]{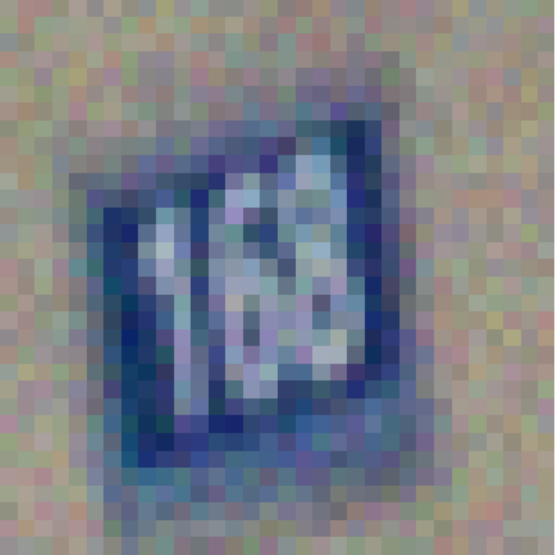}
\caption*{(d)}
\label{fig4d}
\end{minipage}
\caption{Optimized Noise Visualization. The figure illustrates the process of model hijacking where the original task of the model is based on the MNIST dataset, and the model is utilized for a hijacking task using the SVHN dataset. \textbf{(a)} an image from MNIST. \textbf{(b)} an image from SVHN. \textbf{(c)} the optimized noise. \textbf{(d)} an image from SVHN with added noise.}
\label{fig4}
\end{figure}

We define the parameters of the SOL as $\hat{\theta}$, and $\delta$ represents noise that we will add to the hijacking sample to align its data distribution as closely as possible with the original sample, which we will introduce in detail next. Therefore, $f_{\theta}$ and $h_{\hat{\theta}}$ represent the model and layer. To this end, our objective can be formalized as:
\begin{equation}
\begin{aligned}
\hat{\theta} = \arg\min_{\hat{\theta}} \left[\mathcal{L}(\mathcal{D}_{H}\oplus\delta,\mathcal{F})\right]
\end{aligned}
\end{equation}
where $\mathcal{F}=h_{\hat{\theta}}\circ f_{\theta}$, $\mathcal{D}_{H}\oplus\delta=\{x_{i}+\delta | \forall x_{i} \in\mathcal{D}_{H} \}$,
 $\mathcal{L}$ and $\mathcal{D}_{H}$ denoting the loss function and hijacking datasets.

 The model hijacking process leverages the SOL, a component intentionally excluded from the public release to maintain the model's legitimate functionality. Under normal circumstances, the model outputs C1  in response to user queries. However, an adversary equipped with the SOL and knowledge of the model's logits can repurpose the model for a hijacking task.

\subsubsection{Noise Optimization Technique.}
To address C2, which is to ensure the effectiveness of model hijacking attacks under different data distributions, we propose a noise optimization technique, which meticulously calibrates the perturbations introduced into the training dataset. The primary goal of this technique is to adjust the data distribution of the hijacking dataset to be more aligned with that of the original dataset, thereby enhancing the model's ability to generalize and perform well on both datasets. This technique is undergirded by an optimization algorithm that seeks to identify the optimal noise pattern $\delta^{*}$ that minimally perturbs the model's performance on the original task while instilling the capacity to execute the hijacking task. The optimization process of noise must ensure the performance of both the original task and the hijacking task. Thus, we have:
\begin{equation}
\begin{aligned}
&\delta^{*}=\arg\min_{\delta}\left[\mathcal{L}(\mathcal{D}_{M},f_{\theta})+\mathcal{L}(\mathcal{D}_{H}\oplus\delta,\mathcal{F})\right] \\
\end{aligned}
\end{equation}
where $\mathcal{D}_{M}$ denotes the original dataset. The formulation encapsulates the dual objectives of preserving the original task's fidelity and instilling the hijacking task's functionality.

To further elucidate the process of model hijacking, we present in Figure \ref{fig4} a visual representation of the optimized noise that enables the model to perform the hijacking task effectively. 

In the preceding sections and as evidenced by Figure 2d, there is no imperative to extensively obfuscate the physical appearance of the hijacking dataset. This is attributable to the fact that the dataset remains under the adversary's control throughout its lifecycle, thereby rendering unnecessary any attempts to deceive the victim within the physical realm. In the inference phase, the adversary only needs to add the optimized custom noise to the input image when performing the hijacking task.

\subsubsection{Dual-loop Optimization.}
Our proposed solution is dual-loop optimization to tackle C3, balancing the performance between the original task and the hijacking task. Algorithm 1 in the Appendix shows the dual-loop optimization Training. Within the CAMH framework, this approach works by refining the model’s parameters in a synergistic manner, simultaneously enhancing performance for both the original task and the hijacking task. The outer loop focuses on optimizing the original task model parameters $\theta$, while the inner loop fine-tunes the parameters $\hat{\theta}$ for the Synchronized Optimization Layer and the original task model parameters $\theta$, incorporating the optimized noise $\delta^{*}$. This dual-loop mechanism ensures that the model excels in its original function while effectively learning the hijacking task, all with minimal additional computational demand.

\section{Experiment}
In this section, we present our experimental results. We start by introducing our datasets and evaluation settings. Next, we evaluate our CAMH attacks. Finally, we study the impact of some of the hyperparameters in our model hijacking attack.
\subsection{Datasets Description}
We mainly use MNIST~\cite{lecun1998mnist}, SVHN~\cite{SVHN}, GTSRB~\cite{GTSRB}, CIFAR10 \cite{cifar10} and CIFARm datasets for experiments. The specific descriptions of MNIST, SVHN, GTSRB, and CIFAR10 are shown in the Appendix. CIFARm is a dataset we defined ourselves. CIFARm denotes the datasets derived from CIFAR100 \cite{cifar10}. These datasets are formed by randomly selecting $m$ out of the 100 classes available in CIFAR100. The CIFAR100 dataset comprises 100 distinct classes, each class encompassing precisely 600 color images of 32×32 pixel resolution, amounting to a grand total of 60,000 images. By randomly choosing $m$ classes from these 100, we created the CIFARm datasets where $m$ can vary (e.g., 20, 40, 60, 80), simulating tasks of different classification scales. Each CIFARm dataset retains the original image dimensions and color channels of CIFAR100 while reducing the number of classes. This reduction in class count implies less information for models to handle during training and inference. This design allows us to control model complexity and training difficulty while assessing the model's sensitivity to adversarial attacks across varying class sizes.

\subsection{Model Structure and Training Parameters}
To validate the generality and effectiveness of our CAMH method, we conducted experiments using the ResNet18 and ResNet34 \cite{he2016resnet} models, both of which are original models from Torchvision. For ResNet18, we employed 150 training epochs using the SGD optimizer with an initial learning rate of 0.1 and a batch size of 64. Considering the increased complexity and susceptibility to overfitting of the ResNet34 model, we introduced a dropout rate of 0.4 during training and extended the training epochs to 200, other settings remain the same as ResNet18. This configuration ensures the model achieves high accuracy on the original task while providing a stable foundation for subsequent hijacking attack experiments.

\subsection{Evaluation Metrics}
To evaluate the performance of our CAMH attack, we used two metrics, namely the Camouflage Ratio and the Exploitability Ratio.

\textbf{Camouflage Ratio (CR):} The Camouflage Ratio measures the performance of the hijacked model on the original task dataset $\mathcal{D}_{M}$. It is defined as the ratio of the accuracy of the model on $\mathcal{D}_{M}$ after hijacking to the accuracy of the original model on the same $\mathcal{D}_{M}$:
        \begin{equation}
CR = \frac{ACC_{h2o}}{ACC_{o2o}}
\end{equation}
    Here, $ACC_{h2o}$ is the accuracy of the hijacked model on the original task dataset, and $ACC_{o2o}$ is the accuracy of the benign model only trained on the original dataset.

The CR reflects how closely the performance of the hijacked model approaches that of the clean model on a specific original task dataset. As the CR approaches 1, it signifies that hijacked models possess stronger camouflage capabilities, thereby increasing the likelihood of successfully evading detection.

\textbf{Exploitability Ratio (ER) :} The Exploitability Ratio measures the performance of the hijacked model on the hijacking dataset $\mathcal{D}_{H}$. It is defined as the ratio of the accuracy of the hijacked model on $\mathcal{D}_{H}$ to the accuracy of the benign model on the corresponding $\mathcal{D}_{M}$:
    \begin{equation}
ER = \frac{ACC_{h2h}}{ACC_{o2h}}
\end{equation}
    Here, $ACC_{h2h}$ is the accuracy of the hijacked model on the hijacking task dataset, and $ACC_{o2h}$ is the accuracy of the benign model is only trained on hijacked dataset.

The ER evaluates how well the hijacked model maintains its performance when transferred to different datasets. A higher ER indicates greater efficiency in utilization when the hijacker employs the model.

These two metrics collectively form the framework for evaluating the model's performance, enabling a comprehensive understanding of its performance and stability across different environments. In subsequent sections, we will present experimental results based on these metrics and conduct in-depth analyses of the model's performance.

\subsection{The Performance of CAMH}
\begin{figure*}[t]
\centering
\begin{minipage}[t]{0.24\linewidth} 
\centering
\includegraphics[width=\linewidth, height=\linewidth]{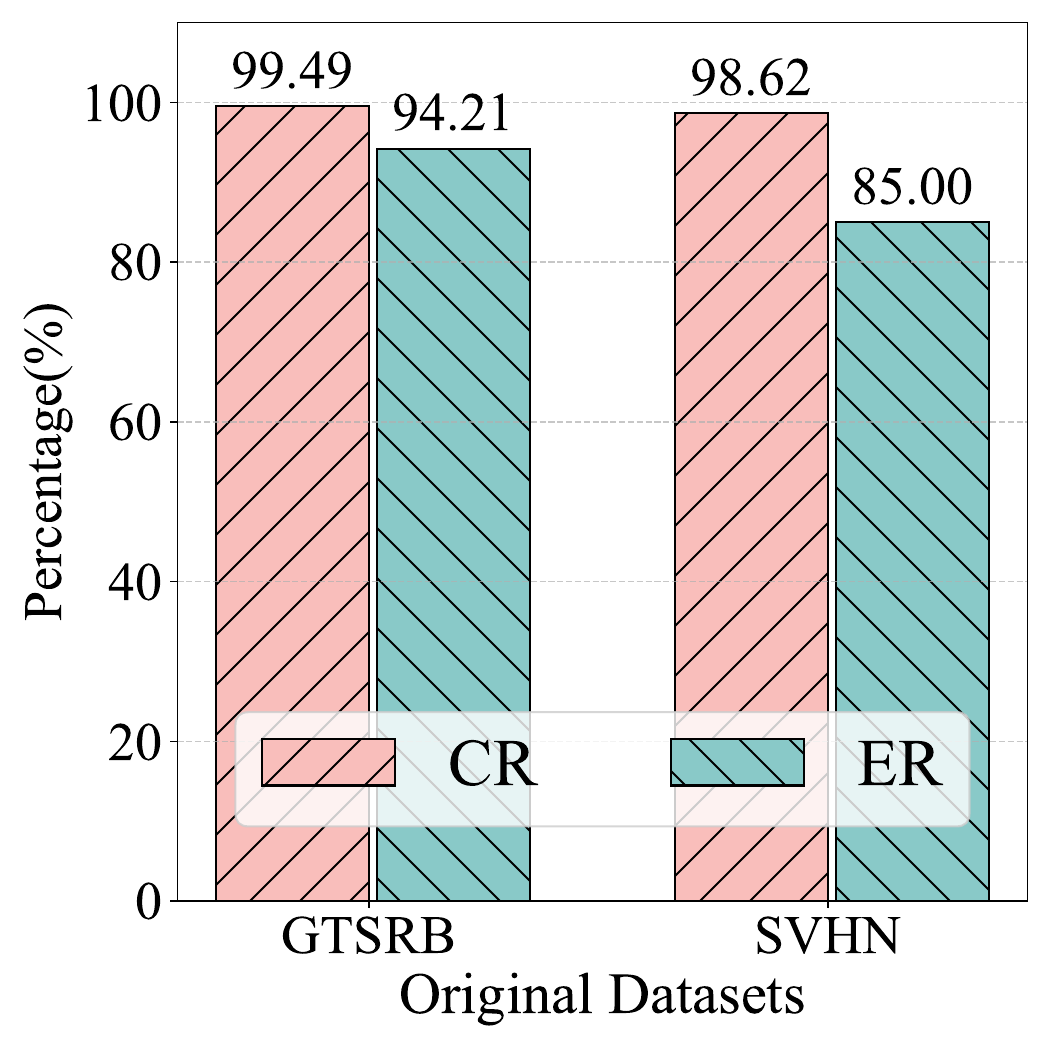} 
\caption*{(a)}
\label{fig2a}
\end{minipage}
\begin{minipage}[t]{0.24\linewidth}
\centering
\includegraphics[width=\linewidth, height=\linewidth]{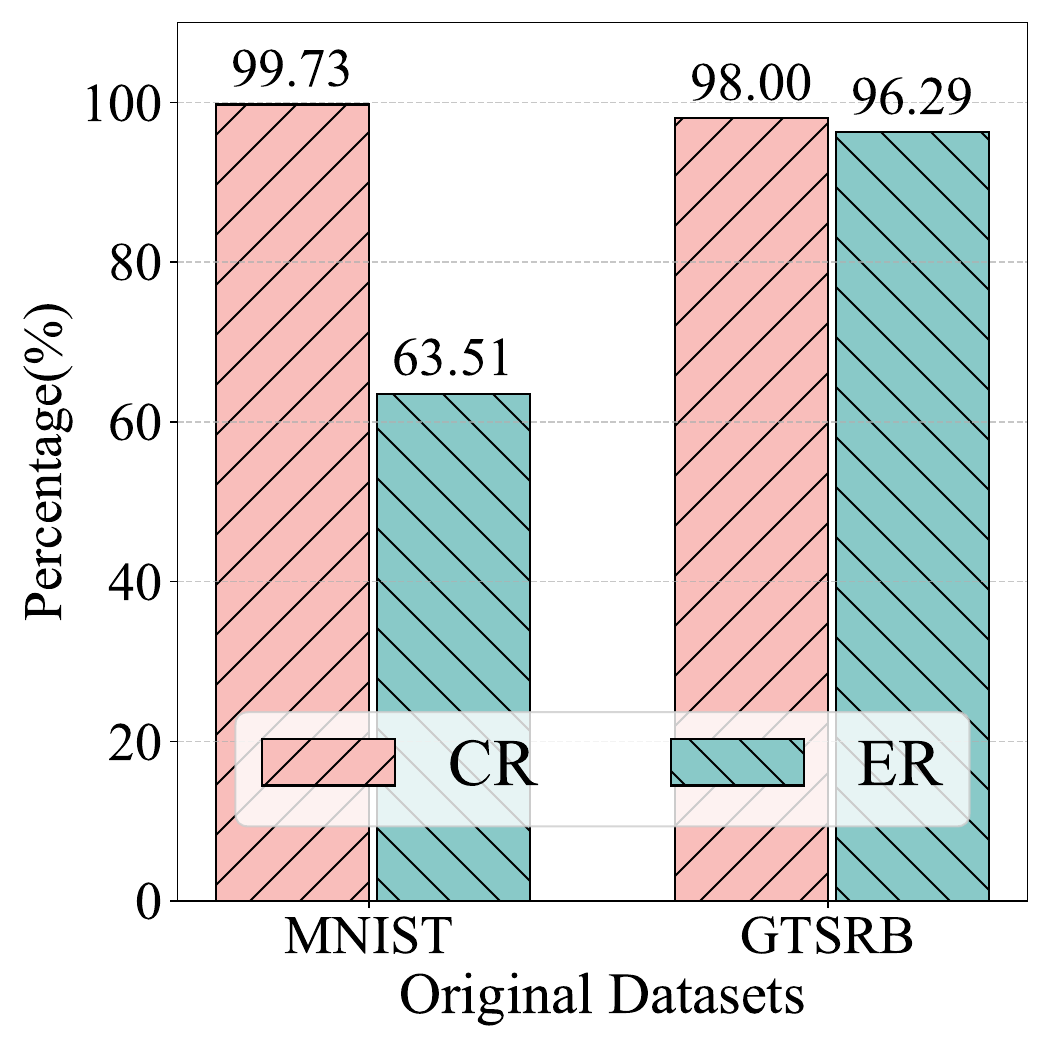}
\caption*{(b)}
\label{fig2b}
\end{minipage}
\begin{minipage}[t]{0.24\linewidth}
\centering
\includegraphics[width=\linewidth, height=\linewidth]{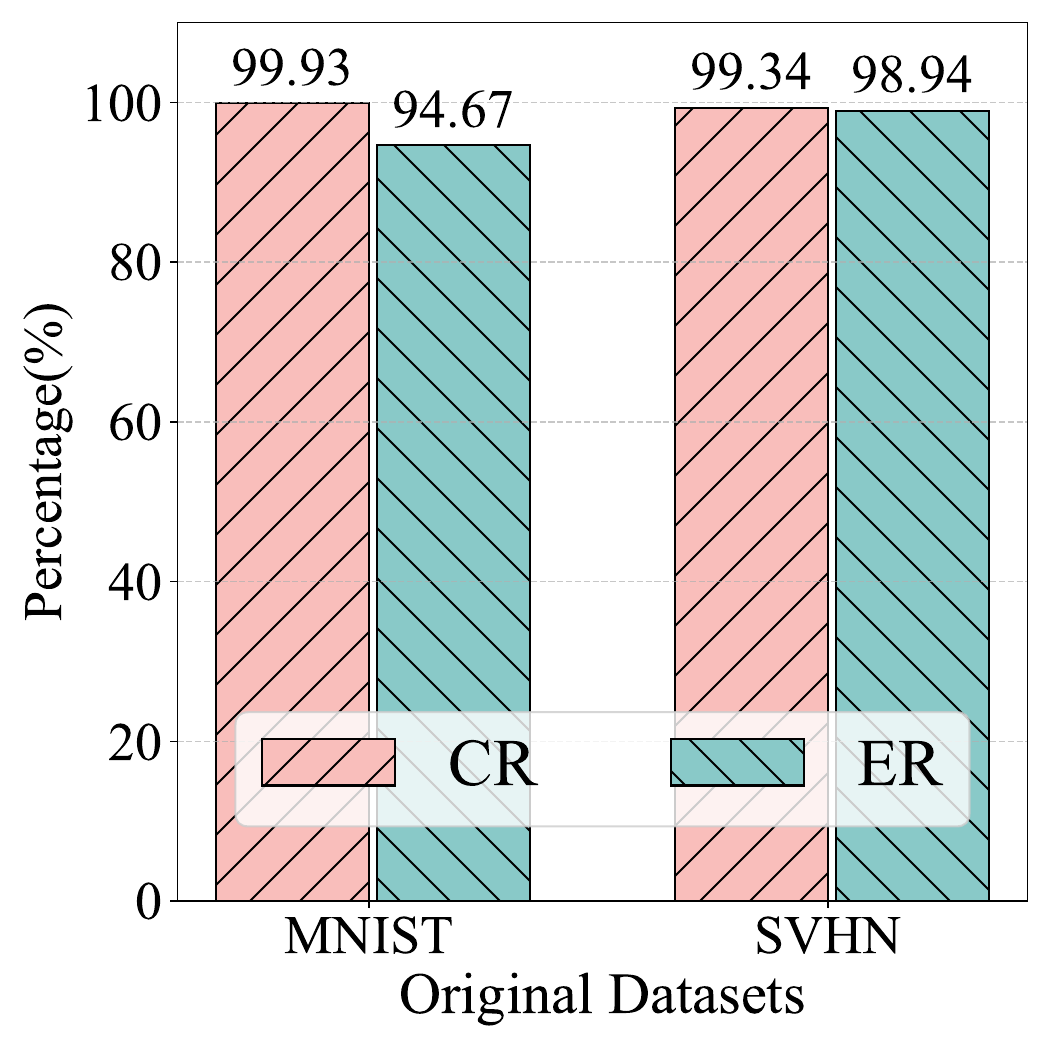}
\caption*{(c)}
\label{fig2c}
\end{minipage}
\begin{minipage}[t]{0.24\linewidth}
\centering
\includegraphics[width=\linewidth, height=\linewidth]{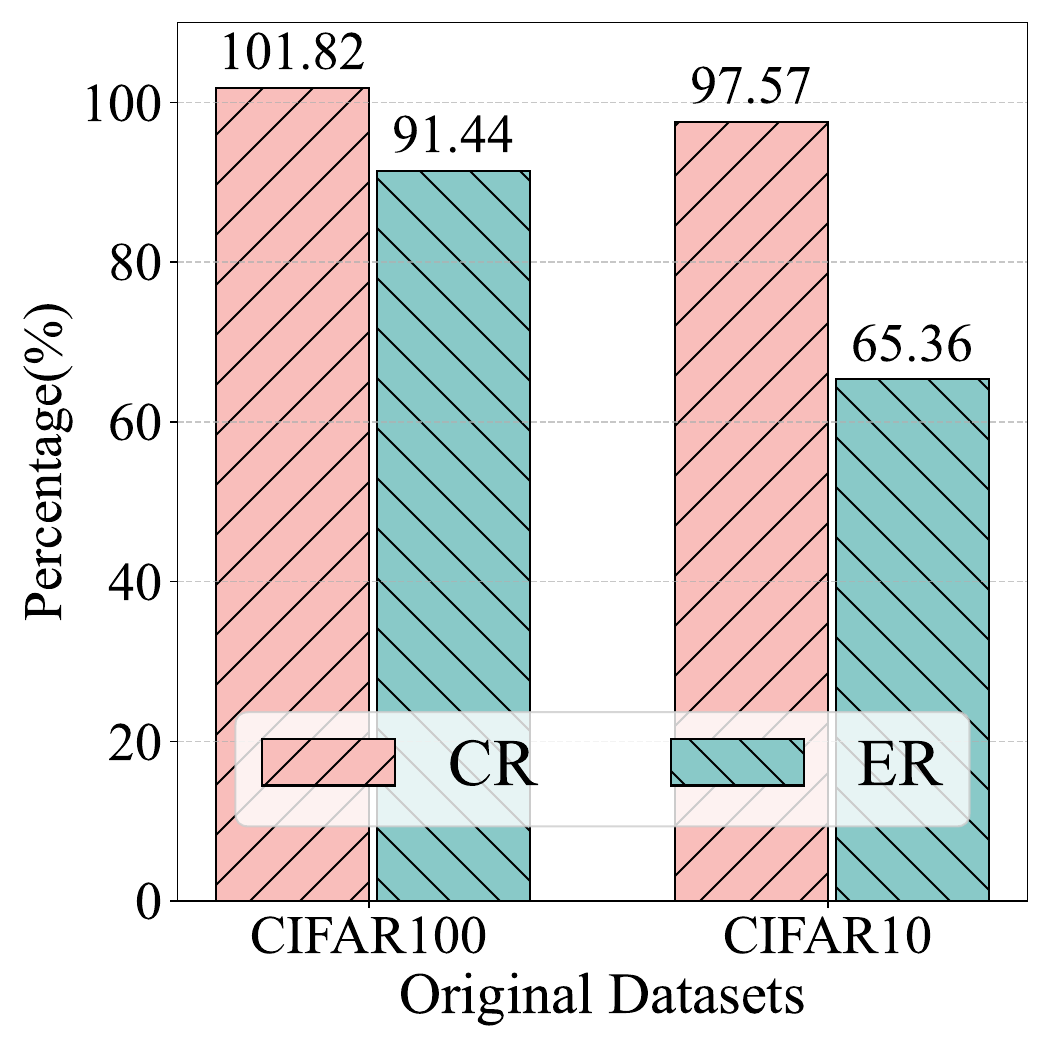}
\caption*{(d)}
\label{fig2d}
\end{minipage}
\begin{minipage}[t]{0.24\linewidth}
\centering
\includegraphics[width=\linewidth, height=\linewidth]{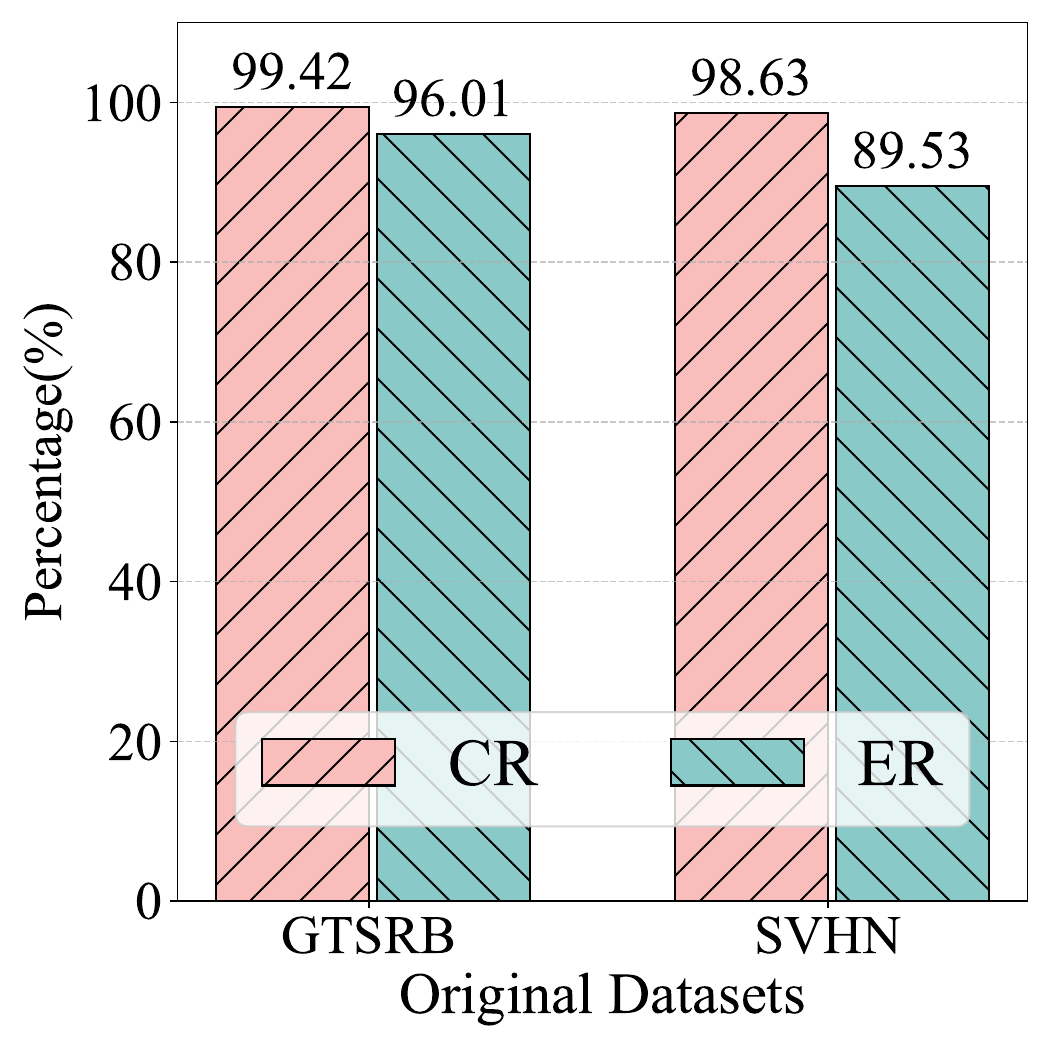}
\caption*{(e)}
\label{fig2e}
\end{minipage}
\begin{minipage}[t]{0.24\linewidth}
\centering
\includegraphics[width=\linewidth, height=\linewidth]{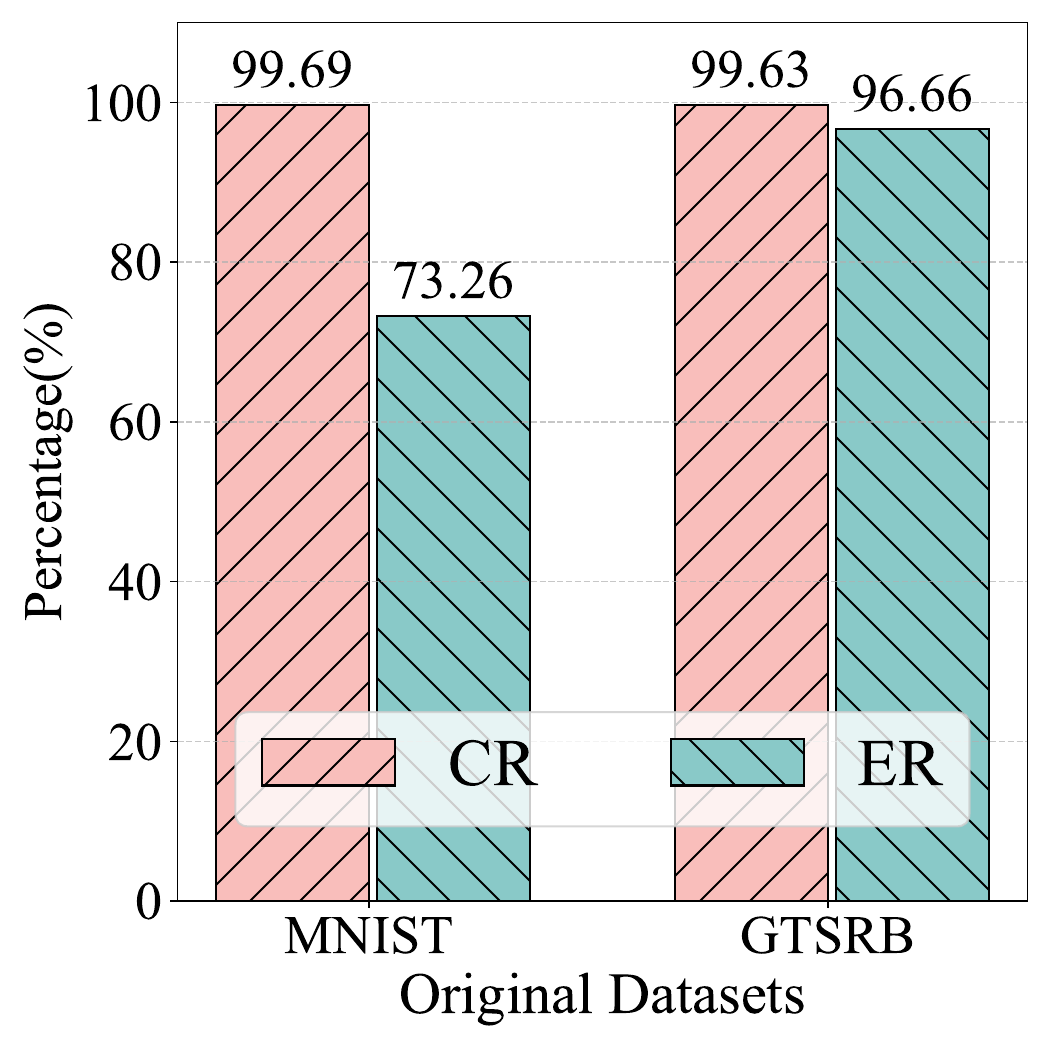}
\caption*{(f)}
\label{fig2f}
\end{minipage}
\begin{minipage}[t]{0.24\linewidth}
\centering
\includegraphics[width=\linewidth, height=\linewidth]{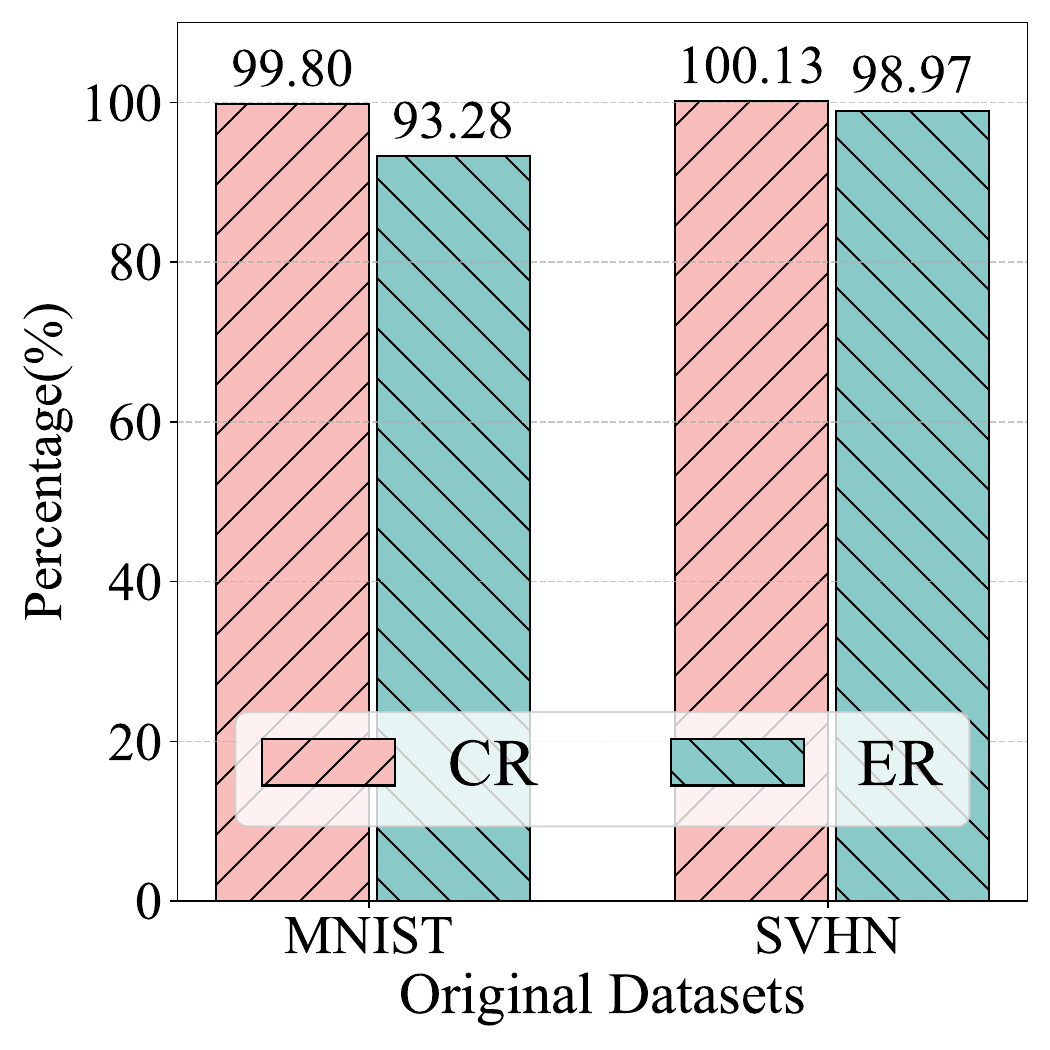}
\caption*{(g)}
\label{fig2g}
\end{minipage}
\begin{minipage}[t]{0.24\linewidth}
\centering
\includegraphics[width=\linewidth, height=\linewidth]{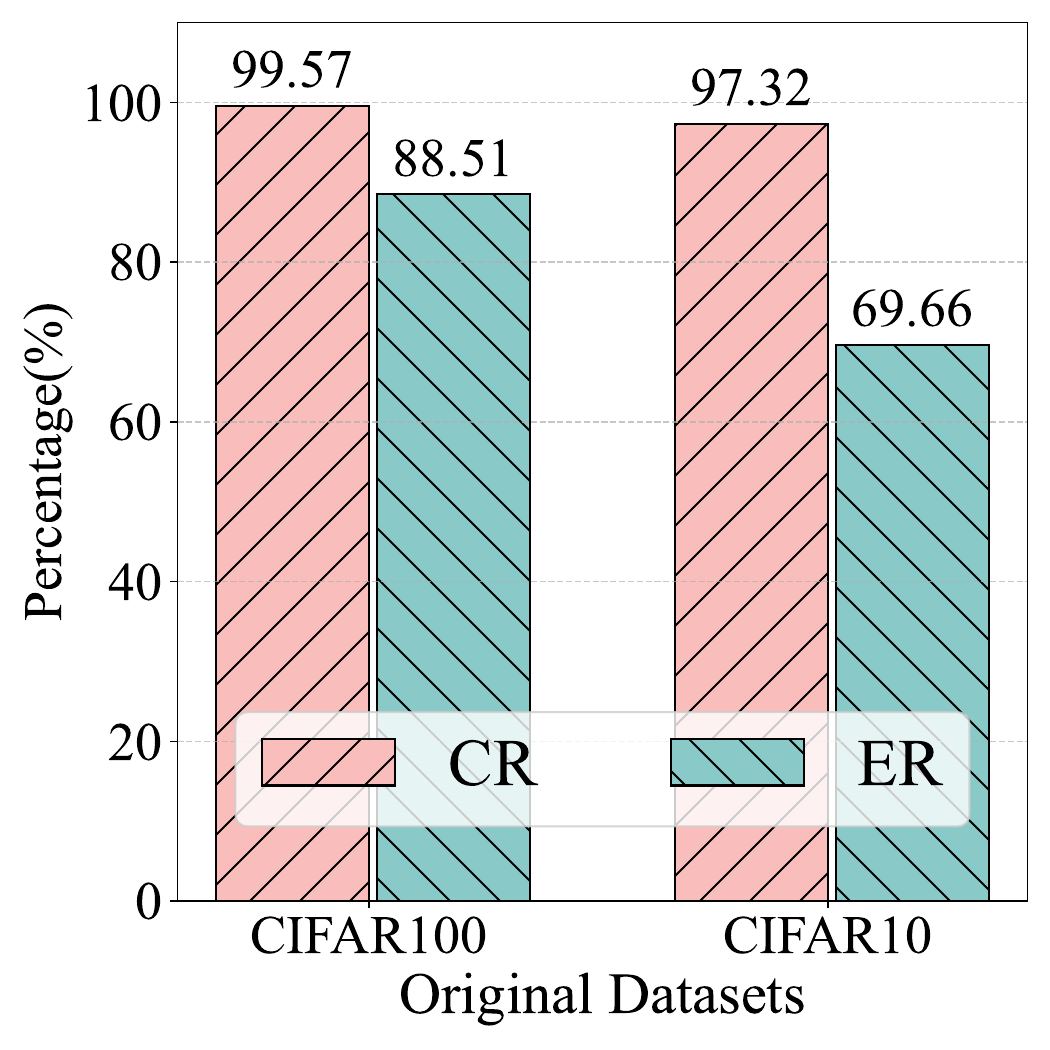}
\caption*{(h)}
\label{fig2h}
\end{minipage}
\caption{The results of our CAMH Attack. The original datasets are denoted on the x-axis, taking Figure \ref{fig2} (a) as an example, when the original task is GTSRB and the hijacking task is SVHN, the model's CR is 99.49\%, and ER is 94.21\%. When the datasets are exchanged, the CR is 98.62\%, and the ER is 85\%. (a)-(d) is for ResNet18, (e)-(h) is for ResNet34. }
\label{fig2}
\end{figure*}

In Figure \ref{fig2}, we present the results demonstrating the CR and ER of various datasets after undergoing CAMH attacks. It is noteworthy that all datasets maintain a CR above around 98\%, indicating that even after model hijacking, their performance on the original tasks remains at a high level. This high CR value increases the likelihood that victims will select our adversarial models from numerous models in the marketplace or that our models will pass the accuracy acceptance criteria of the first party more easily. In some cases, the CR exceeds 100\%. This occurs because the inclusion of the hijacked dataset enhances the model's generalization ability on the original task, especially when the data distributions of the two datasets are not significantly different.

\begin{figure}[t]
\centering
\includegraphics[width=\linewidth, height=0.6\linewidth]{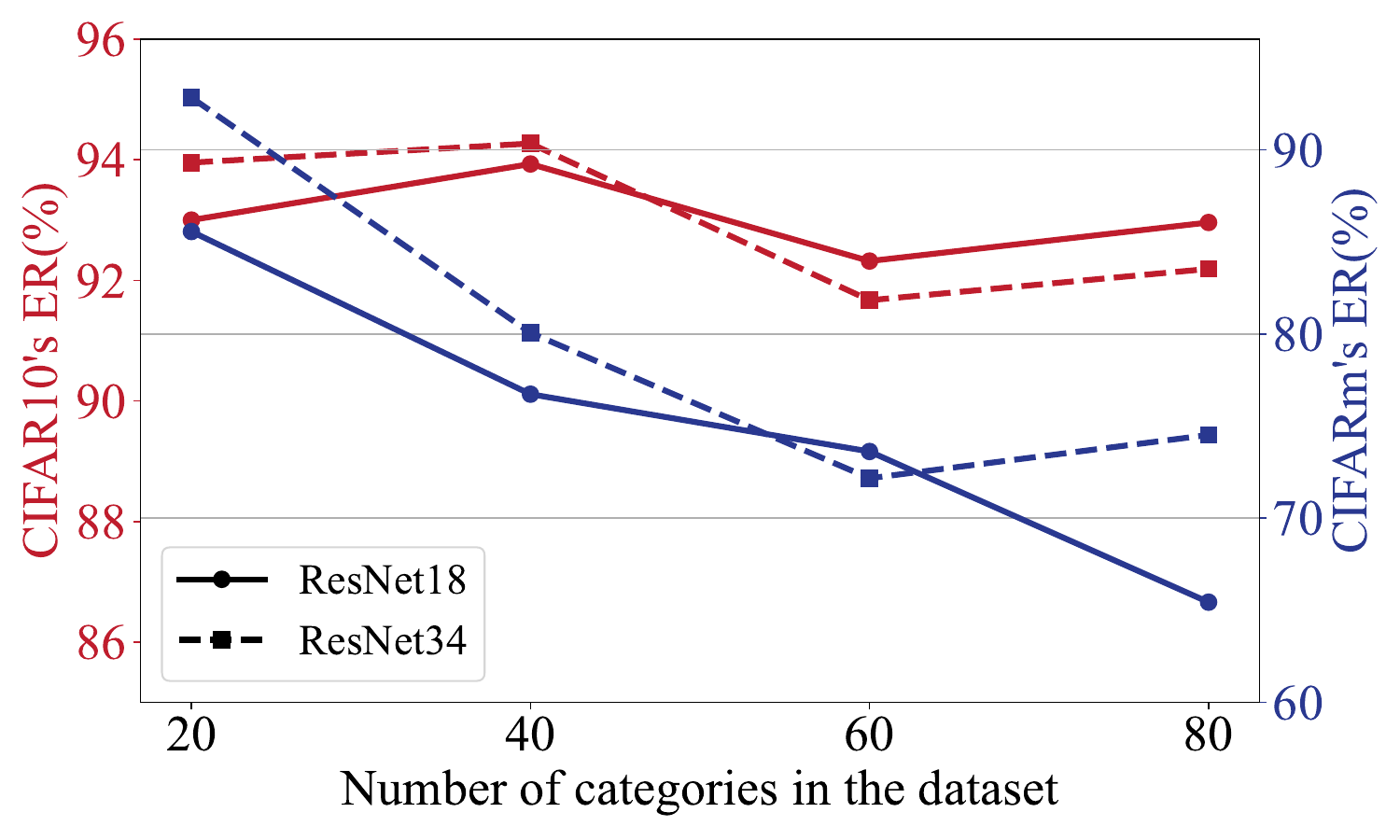} 
\caption{The impact of the number of categories on the ER of hijacking tasks. The red line illustrates the relationship between the ER of the hijacking task and the value of m when the original task is CIFARm, the hijacking task is CIFAR10, and the CR remains above 95\%. The blue line represents the situation when the original and hijacking tasks are swapped. We compared ResNet18 and ResNet34.}
\label{fig3}
\end{figure}

Simultaneously, ER values indicate the efficiency of the models in performing hijacking tasks, with most datasets achieving an ER of over 85\%. However, in cases where the original dataset is MNIST and the hijacking dataset is GTSRB, or where the original task is CIFAR100, and the hijacking task is CIFAR10, ER values show poorer performance but still exceed 60\%. This may be due to the hijacking dataset having a larger number of categories than the original dataset, with significant differences in data distribution that exceed the capability of the hijacking method.

For instance, in Figure \ref{fig2}b, the original dataset has fewer categories (10) compared to the hijacking dataset (43), and there is substantial dissimilarity between MNIST and GTSRB. The combination of these factors results in ER values lower than anticipated. If only one condition is met, as shown in Figure \ref{fig2}a or \ref{fig2}c, the execution of hijacking tasks still falls within CAMH's hijacking capability range.

Let's focus on the second and fourth columns. We observe that more complex models exhibit stronger transferability, especially in challenging tasks. For the original task MNIST and the transfer task GTSRB, ResNet34 shows an increase of nearly 10\% in ER compared to ResNet18. Similarly, for the original task CIFAR10 and the transfer task CIFAR100, ResNet34 demonstrates a 4\% increase in error rate over ResNet18. This phenomenon may be attributed to the more complex models having a greater number of parameters and deeper network structures, allowing them to learn richer feature representations.

\subsection{The Impact of the Number of Categories}
In this section, we investigate the impact of the number of categories on model hijacking capability. As illustrated in Figure \ref{fig3}, we examine the performance of the original task and the hijacking task across datasets with varying numbers of categories. While maintaining the baseline performance of the original task, we find that the model's hijacking capability excels when the hijacking task faces fewer categories than the original task, achieving ER values exceeding 90\% for both ResNet18 and ResNet34. Conversely, as the number of categories in the adversarial task exceeds that of the original task, ER values decrease, irrespective of whether ResNet18 or ResNet34 is employed. We observe that in such scenarios, ResNet34 outperforms ResNet18, suggesting that larger-scale models may handle more challenging hijacking tasks with greater ease.

\subsection{The Impact of Hijacking Data Volume}
\begin{figure}[t]
\centering
\includegraphics[width=\linewidth, height=0.8\linewidth]{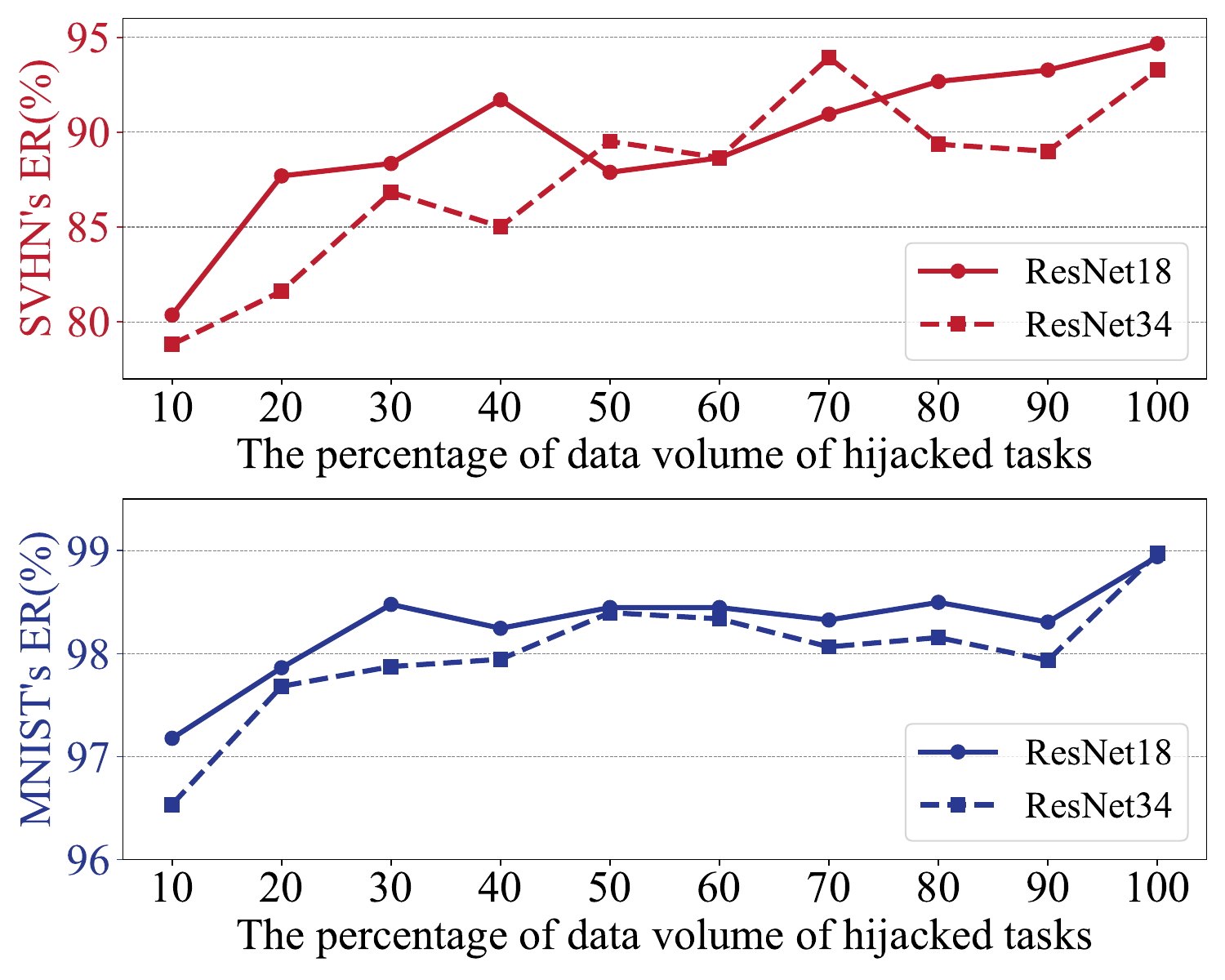} 
\caption{The impact of the percentage of data volume of hijacking tasks on the ER of hijacking tasks. The red line illustrates the relationship between the ER of the hijacking task and the value of m when the original task is MNIST, the hijacking task is SVHN, and the CR remains above 95\%. The blue line represents the situation when the original and hijacking tasks are swapped. We compared ResNet18 and ResNet34. }
\label{fig5}
\end{figure}

In this section, we investigate the impact of hijacking data volume on model hijacking tasks. Figure \ref{fig5} vividly demonstrates that the scale of hijacking data significantly influences the ER of model hijacking tasks. Specifically, as the volume of data increases, both curves exhibit an upward trend. When the volume of hijacking data constitutes only 30\% of the total, the model's ER reaches a satisfactory high level. This phenomenon is particularly pronounced in the SVHN and MNIST datasets: SVHN achieves an ER exceeding 85\%, while MNIST approaches an ER over 98\%.

This finding is of profound significance as it indicates that even with a relatively modest volume of hijacking data, our model hijacking method achieves efficient task execution. This efficacy is clearly delineated in Figure \ref{fig5}, where the red and blue lines denote the ER variations of the SVHN and MNIST datasets across varying percentages of hijacking data volumes, respectively.

Furthermore, we observe that as the volume of hijacking data continues to increase, the ER improves incrementally, albeit at a diminishing rate. This suggests the presence of an optimal data volume range beyond which the marginal benefit of increased data diminishes in enhancing hijacking task performance.

Our findings imply that even with constrained data volumes, adversaries may proficiently manipulate models to execute illicit tasks via meticulously model hijacking. 

\subsection{Ablation Study}

\begin{table}[ht]
\centering
\begin{tabular}{c|c|c|c|c}
\toprule
Model & Noise & SOL & CR & ER \\
\midrule
\multirow{4}{*}{ResNet18} & \texttimes & \texttimes & 99.89\% & 92.23\% \\
 & \checkmark & \texttimes & 99.95\% & 93.60\% \\
 & \texttimes & \checkmark & 99.95\% & 91.97\% \\
 & \checkmark & \checkmark & \textbf{99.93}\% & \textbf{94.67\%} \\
\midrule
\multirow{4}{*}{ResNet34} & \texttimes & \texttimes & 99.84\% & 88.37\% \\
 & \checkmark & \texttimes & 99.79\% & 91.55\% \\
 & \texttimes & \checkmark & 99.79\% & 91.95\% \\
 & \checkmark & \checkmark & \textbf{99.80}\% & \textbf{93.28\%} \\
\bottomrule
\end{tabular}
\caption{The results of the ablation study with MNIST as the original task and SVHN as the hijacking task. }
\label{tab:ablation}
\end{table}
In Section 3.3, we delineate our approach to optimizing a noise and layer-wise projection mapping. To assess the impact of noise and SOL-based projections, we have undertaken ablation study.

To incorporate the layer into our analysis, we have selected MNIST as the baseline task and SVHN as the target for model hijacking. The findings detailed in Table \ref{tab:ablation} indicate that when the attacker employs only the optimized noise, the ER for the hijacking task escalates. When the synchronized layer is used exclusively, there is an improvement in performance. The hijacking performance of the model is notably enhanced when both strategies are combined, culminating in the highest ER. It is noteworthy that, due to the consistently high CR, our impact on the original task remains minimal across all scenarios, demonstrating strong camouflage capabilities.

\section{Conclusion}
In this paper, we introduce a Category-Agnostic Model Hijacking (CAMH) attack for the first time. The CAMH method effectively solves the key challenges in model hijacking through innovative multi-category hijacking task design, noise optimization technology, and double-loop optimization. Experimental results show that CAMH can maintain the high performance of the original task while successfully performing the hijacking task even when the number of categories does not match and the data distribution is significantly different. In the future, our research will focus on enhancing detection and defense mechanisms against such attacks and expanding the exploration of the application and impact of model hijacking across various scenarios.

\section*{Acknowledgment}
This work was partly supported by the National Key Research and Development Program of China under No. 2022YFB3102100, NSFC under No. U244120033, U24A20336, 62172243 and 62402425, and the Zhejiang Provincial Natural Science Foundation under No. LD24F020002.

\bigskip
\bibliography{aaai25}

\clearpage
\newpage

\newcommand{\appendixtitle}[1]{
\vspace*{0cm}
\begin{center}
{\huge #1}
\end{center}
\vspace{0.2cm}
}

\renewcommand{\thesection}{\Alph{section}} 
\setcounter{section}{0}

\appendixtitle{Appendix}

\section{Hardware and Environment}
In summary, our experiments were conducted utilizing the PyTorch framework for implementation. We executed all experiments on an NVIDIA GeForce RTX 3090 GPU, which is equipped with 24GB of dedicated memory. The software environment consisted of CUDA version 11.7, PyTorch version 2.0.1, and Ubuntu version 20.04.1, ensuring an optimal configuration for our computational tasks.

\section{Dual-loop Optimization Training}
\begin{algorithm}[h]
\caption{Dual-loop Optimization Training}
\label{alg:algorithm}
\textbf{Input}: Hijacking dataset $\mathcal{D}_{H}$, Original task dataset $\mathcal{D}_{M}$, \\
\textbf{Output}: Perturbation $\delta$, Model $f_{\theta}$, SOL $h_{\hat{\theta}}$

\begin{algorithmic}[1] 
\STATE Initialization: $\delta\leftarrow 0$.
\FOR{epoch in epochs}
\STATE $\theta = \arg\min_{\theta} \mathbb{E}_{(x_i,y_i)\in\mathcal{D}_{M}}\left[\mathcal{L}(x_i,y_i;f_{\theta})\right]$ \\
\STATE // Outer loop optimizing the original task model parameters
\STATE $\hat{\theta} = \arg\min_{\hat{\theta}} \mathbb{E}_{(x_i,y_i)\in\mathcal{D}_{H}}\left[\mathcal{L}(x_i\oplus\delta,y_i;h_{\hat{\theta}}\circ f_{\theta})\right]$ \\
\STATE // Inner loop optimizing the parameters for the SOL
\STATE $\delta = \arg\min_{\delta} \mathbb{E}_{(x_i,y_i)\in\mathcal{D}_{H}}\left[\mathcal{L}(x_i\oplus\delta,y_i;h_{\hat{\theta}}\circ f_{\theta})\right]$ \\
\STATE //Inner loop synchronous optimizing noise

\ENDFOR
\STATE \textbf{return} $\delta$,$f_{\theta}$ and $h_{\hat{\theta}}$
\end{algorithmic}
\end{algorithm}

\section{Datasets Description}
\begin{itemize}
\item \textbf{MNIST} is a grey-scale handwritten digits classification dataset. It consists of 70,000 images, each of them in the size of 28×28. The MNIST dataset is equally split between 10 classes.
\item \textbf{SVHN} (Street View House Numbers) dataset is a widely used image recognition dataset composed of digits cropped from street view images, uniformly scaled to 32×32 pixels. This dataset comprises ten categories corresponding to digits 0 through 9. 
\item \textbf{GTSRB} dataset is widely recognized as a benchmark for traffic sign recognition. It comprises over 40 categories, including but not limited to speed limits, stop signs, yield signs, traffic lights, and more. With a total of over 50,000 real-world traffic sign images, ranging in size from 15×15 to 250×250 pixels, the dataset offers rich diversity and real-world scenarios.
\item \textbf{CIFAR10} is a 10 classes colored dataset. It consists of 60,000 images with a size of 32×32. The images are equally split between the following ten classes. 
\end{itemize}

\section{Original Model Capability}
Since our evaluation metrics require the capability of the benign model on the original datasets, we have conducted experiments for each benign model on all datasets separately. As shown in Table \ref{table1}, in this way, we can understand the specific impact our method will have on the model. We can observe that, under the current settings, except for the GTSRB dataset, ResNet34 performs better on individual datasets than ResNet18, which benefits from a more complex model structure and a greater number of network parameters.
\begin{table}[h]
\centering
\begin{tabular}{c|c|c|c}
    \toprule
    \multirow{2}{*}{Dataset}  & \multicolumn{3}{c}{Model}  \\
    \cmidrule{2-4}
          & ResNet18 & ResNet34 & DenseNet121\\
    \midrule
    MNIST & 99.22\% & 99.27\% & 99.33\% \\
    SVHN  & 95.52\% & 95.64\% & 95.63\% \\
    GTSRB  & 92.71\% & 91.35\% & 94.44\% \\
    CIFAR10  & 88.18\% & 88.46\% & 89.54\% \\
    CIFAR100  & 61.57\% & 62.07\% & 65.64\% \\
    CIFAR20   & 75.29\% & 76.38\% & 77.43\% \\
    CIFAR40   & 72.47\% & 72.51\% & 75.43\% \\
    CIFAR60  &  67.39\%  & 68.13\% & 69.33\% \\
    CIFAR80   &  63.26\%  & 64.25\% & 67.83\% \\
    \bottomrule
\end{tabular}
\caption{The performance of benign model on original datasets.}
\label{table1}
\end{table}

\section{Comparison with Baseline}
\subsection{Setting of Comparison Method}
We conducted the experimental setup in accordance with the requirements specified in the paper. Due to the limitation that the number of hijacked dataset categories in The Chameleon Attack and The Adverse Chameleon Attack must not exceed the number of categories in the original dataset, we selected the SVHN and MNIST datasets for comparison with our CAMH. We chose 10,000 samples for the camouflager training, with 50 training epochs, and the feature extractor was set to MobileNetV2. After the completion of the camouflager training, 20000 disguised samples were generated by the camouflager to poison the original dataset. Finally, the poisoned dataset was used to hijack the ResNet18, with 100 training epochs and the loss function being CrossEntropyLoss. All batch sizes were set to 64, the optimizers were Adam, and the learning rate was 0.001.
\subsection{Comparison Results}
\begin{figure}[t]
\centering
\includegraphics[width=0.9\columnwidth]{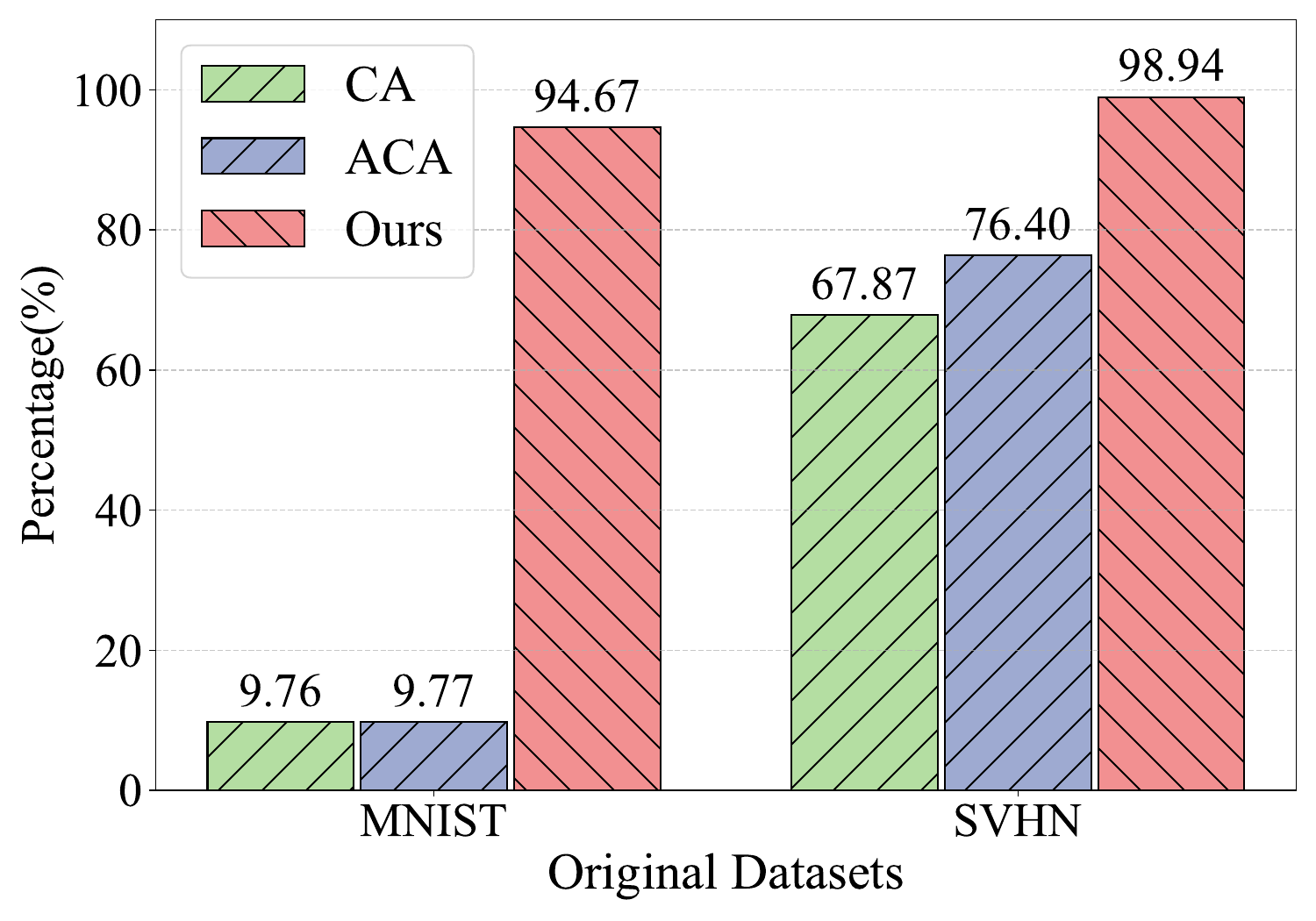} 
\caption{ER values of CAMH attack, The Chameleon attack, and The Adverse Chameleon attack.}
\label{fig6}
\end{figure}

We use CA to represent The Chameleon Attack and ACA to represent The Adverse Chameleon Attack. Since the code for the comparison methods was not open-sourced, we reproduced the results based on the descriptions in the respective papers. As shown in Figure \ref{fig6}, when the original dataset is MNIST, and the hijacking dataset is SVHN, both CA and ACA perform poorly in executing the hijacking attack, with ER values below 10\%. However, when the original dataset is SVHN and the hijacking dataset is MNIST, CA and ACA perform better. Regardless of the scenario, CAMH consistently achieves an ER value of over 94\%. In addition, all CR values are close to 1, so we don’t show them in the figure.

\section{Defence Evaluation}
We referred to defense methods from previous studies. A possible defense approach is to employ an autoencoder or various denoising techniques on both training and testing images. We conducted defense tests on Resnet-34 with GTSRB as the original task and SVHN as the hijacking task. The following are the experimental results obtained by using different denoising methods.
\begin{table}[h]
\centering
\begin{tabular}{c|c|c|c|c}
\midrule
Method & GTSRB(acc) & SVHN(acc) & CR & ER \\
\midrule
Nodefence & 90.82 & 91.82 & 99.42 & 96.01 \\
Wavelet & 87.43 & 91.15 & 95.71 & 95.31 \\
Median & 88.81 & 90.10 & 97.22 & 94.21 \\
Bilateral & 30.62 & 71.17 & 33.52 & 74.41 \\
Autoencoder & 35.65 & 69.75 & 39.03 & 72.93 \\
\midrule
\end{tabular}
\caption{Experimental results of different defence methods}
\label{table2}
\end{table}

As shown in Table \ref{table2}, we can see that the first two methods are not effective in reducing the hijacking effect. While the last two methods reduce the ER value, they have a significant negative impact on the original task performance. We plan to further explore different defense techniques that may be more effective.

\section{More models}
\begin{figure}[t]
\centering
\begin{minipage}[t]{0.48\linewidth} 
\centering
\includegraphics[width=\linewidth, height=\linewidth]{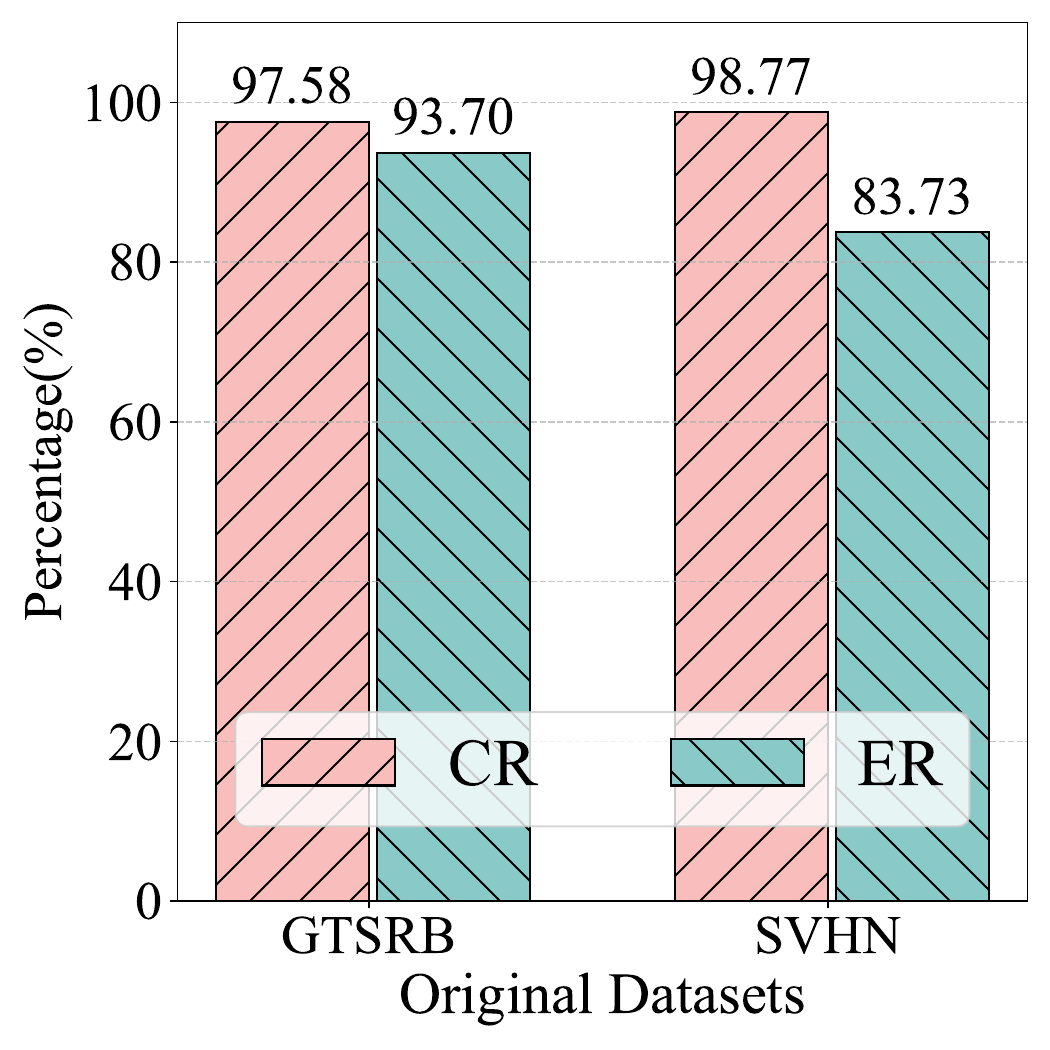} 
\caption*{(a)}
\label{fig7a}
\end{minipage}
\begin{minipage}[t]{0.48\linewidth}
\centering
\includegraphics[width=\linewidth, height=\linewidth]{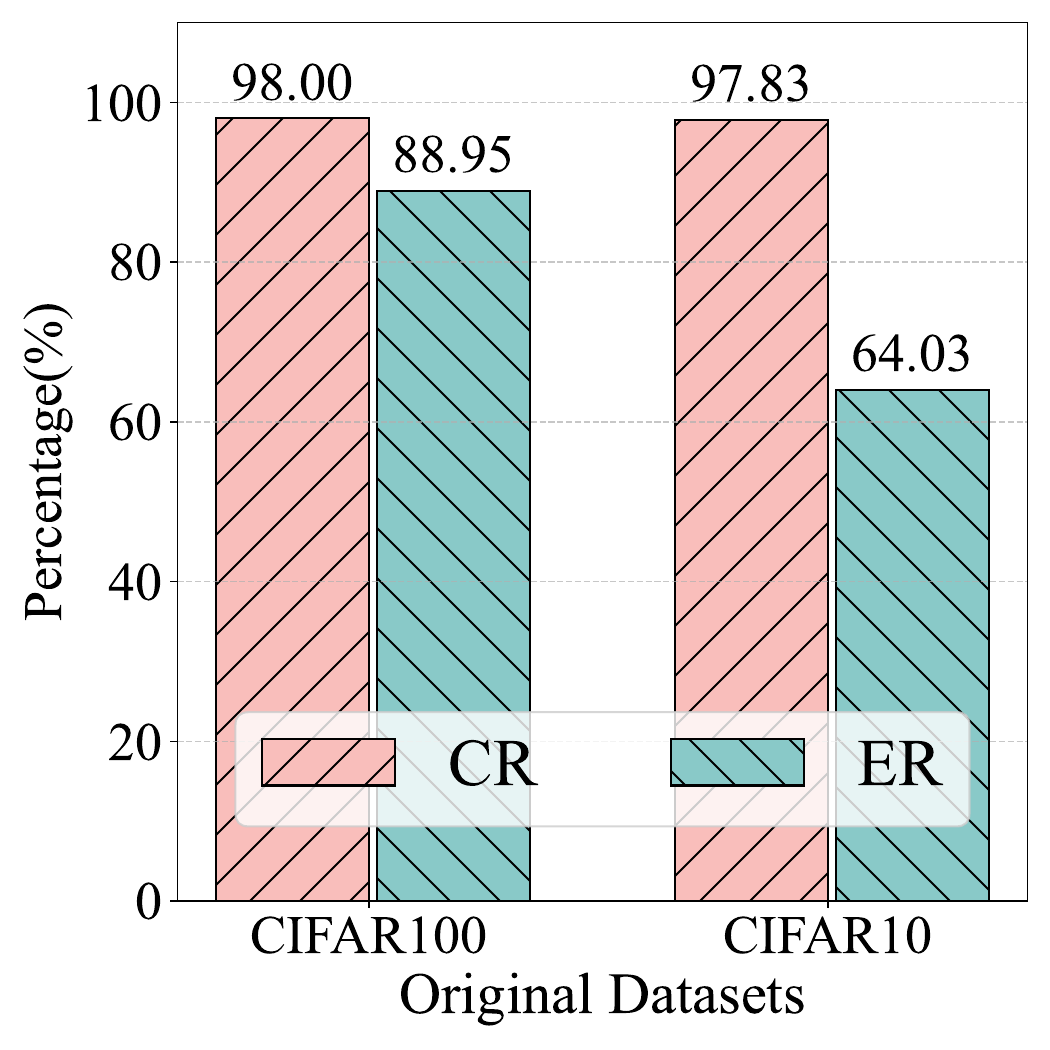}
\caption*{(d)}
\label{fig7d}
\end{minipage}
\caption{The effect of CAMH on DenseNet-121 }
\label{fig7}
\end{figure}

We additionally carried out corresponding experimental studies on the DenseNet-121 model. During this experiment, all the parameters were strictly set in accordance with those previously configured, and the number of training epochs was still maintained at 150. As the figure \ref{fig7} shows, it can be found that all the evaluation indicators are at a relatively ideal level. This indicates that under the DenseNet-121 model, adopting the same parameters and training settings can also achieve satisfactory results, further verifying the effectiveness and universality of the methods we adopted.

\end{document}